\RequirePackage{fix-cm}
\documentclass[twocolumn]{svjour3}          
\smartqed  
\usepackage{graphicx}
\usepackage{natbib}
\journalname{Rendiconti Lincei. Scienze Fisiche e Naturali. LINCEI PRIZEWINNERS}
\begin{document}

\title{The ``dynamical clock'': dating the internal dynamical
  evolution of star clusters with Blue Straggler Stars\thanks{The
    author F. R. Ferraro received the international prize
    ``L. Tartufari'' for Astronomy in 2018, attributed by the
    Accademia Nazionale dei Lincei, Rome.}  }

\titlerunning{The stellar dynamical clock}        

\author{Francesco R. Ferraro \and
  Barbara Lanzoni \and
  Emanuele Dalessandro
}


\institute{Francesco R. Ferraro \at
           Dipartimento di Fisica e Astronomia, Universit\`a degli Studi di Bologna, Via Gobetti 93/2, I--40129 Bologna, Italy \\
              Tel.: +39-051-2095774\\
              \email{rancesco.ferraro3@unibo.it}           
           \and
           Barbara Lanzoni \at
           Dipartimento di Fisica e Astronomia, Universit\`a degli Studi di Bologna, Via Gobetti 93/2, I--40129 Bologna, Italy 
           \and
           Emanuele Dalessandro \at
           INAF-Osservatorio di Astrofisica \& Scienza dello Spazio, via Gobetti 93/3, I--40129 Bologna, Italy
}

\date{Received: November 12, 2019 / Accepted: January 12, 2020}

\maketitle

\begin{abstract}
We discuss the observational properties of a special class of objects
(the so-called ``Blue Straggler Stars'', BSSs) in the framework of
using this stellar population as probe of the dynamical processes
occurring in high-density stellar systems. Indeed, the shape of the
BSS radial distribution and their level of central concentration are
powerful tracers of the stage of dynamical evolution reached by the
host cluster since formation. Hence, they can be used as empirical
chronometers able to measure the dynamical age of stellar systems. In
addition, the presence of a double BSS sequence in the color-magnitude
diagram is likely the signature of the most extreme dynamical process
occurring in globular cluster life: the core collapse event. Such a
feature can therefore be used to reveal the occurrence of this process
and, for the first time, even date it.
\end{abstract}

\section{Introduction}
\label{sec_intro}
\begin{figure*}
\includegraphics[scale=0.8]{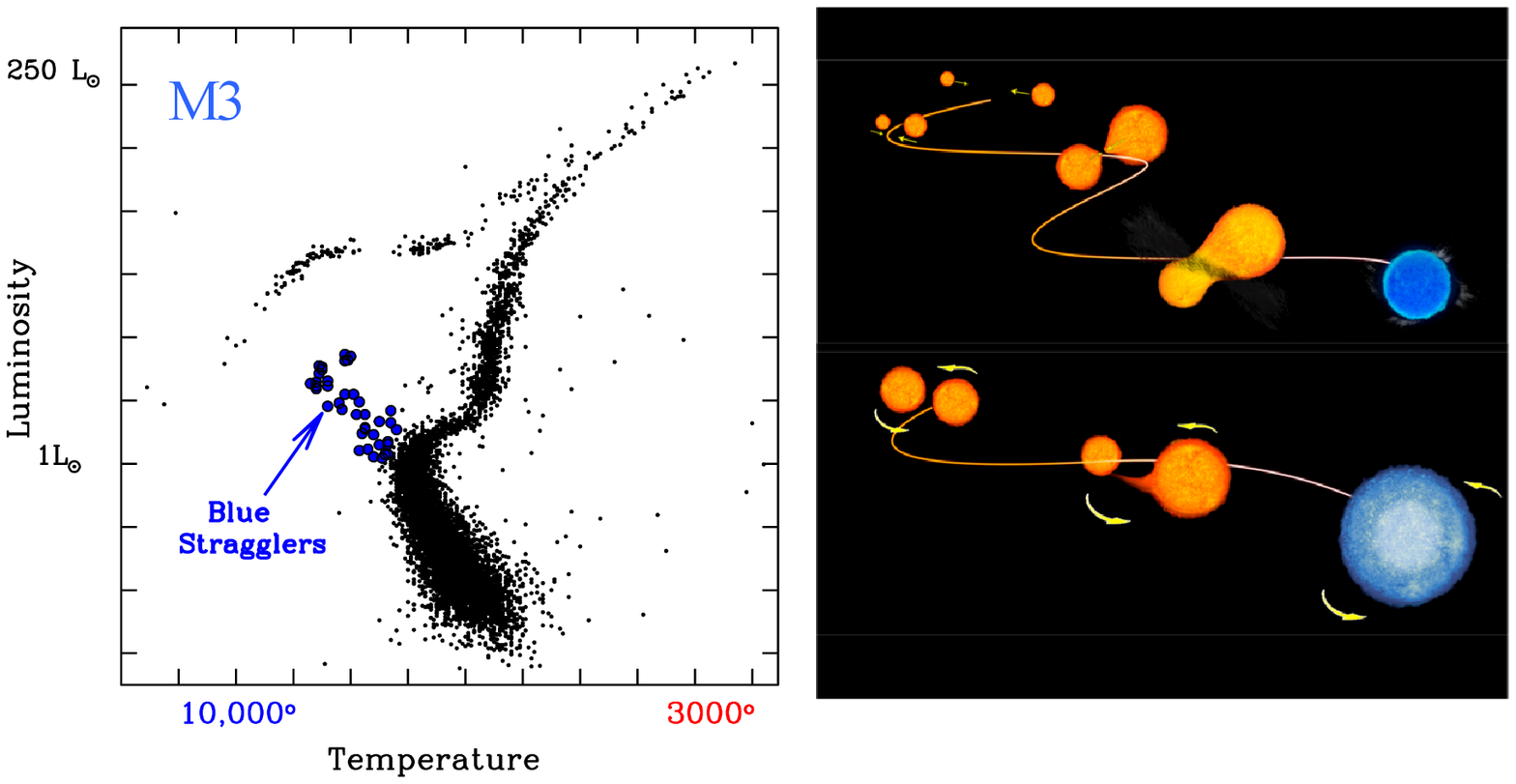}
\caption{Left: indicative illustration of the location of BSSs (large
  blue circles) in a Temperature - Luminosity diagram.  Luminosities
  are expressed in Solar units, temperatures in Kelvin. Right:
  artistic illustration of the two main BSS formation channels, namely
  stellar collisions (upper panel), and vampirism phenomena between
  two companion stars in a binary system (lower panel).}
\label{fig:1}       
\end{figure*}

Globular Clusters (GCs) are compact aggregates of up to a million
stars held together by their mutual gravitational attraction in a
nearly spherical configuration. They are sub-galactic structures, with
typical masses of $10^4-10^6 M_\odot$, ages as old as the Hubble time
($\sim 13$ Gyr), and cores that present the most extreme stellar
densities in the Universe, reaching up to a few millions of stars per
cubic parsec. GCs are the most populous and oldest systems where stars
can be individually observed.  At odds with what happens in galaxies,
where the orbital motion of stars primarily depends on the average
gravitational potential, GCs are ``collisional'' systems, where
two-body interactions cause kinetic-energy exchanges among stars and
gravitational perturbations to their orbits, bringing the cluster
toward a thermodynamically relaxed state in a timescale (relaxation
time) that can be significantly shorter that its age. For this reason
they represent unique cosmic laboratories to study the fundamental
physical processes characterizing multi-body dynamics. Because of such
interactions, heavy stars tend to progressively sink toward the
central region of the cluster (dynamical friction), while low-mass
stars can escape from the system (evaporation). This yields a
progressive contraction of the core, producing an impetuous increase
of its density virtually toward infinity: the so-called ``core
collapse''. The runaway contraction is thought to be halted by the
formation and hardening of binary systems, and the post-core collapse
phase is characterized by core oscillations, with several episodes of
high central density followed by stages during which the cluster
rebounds toward a structure with lower density and more extended
core. The recurrent gravitational interactions among stars thus modify
the structure of the system over the time (the so-called "dynamical
evolution"), with a time-scale (the relaxation time) that depends in a
very complex way on the initial and the local conditions, thus
differing from cluster to cluster and, within the same system, from
high- to low-density regions (e.g., \citealt{meylan+97}).

As a consequence of the internal dynamical evolution, clusters born
with a given size progressively develop more and more compact cores,
the timescale of these changes being hard to determine. Indeed,
estimating the formation epoch of a cluster (corresponding to the
chronological age of its stars) is relatively simple from the measure
of the luminosity of the Main Sequence Turn-Off (MS-TO) level, while
measuring its ``dynamical age'' (corresponding to the level of
dynamical evolution it reached since formation) is much more
challenging. Following an analogy to human experience, as people with
the same biological age can be in very different physical shapes, so
stellar aggregates with the same chronological age can have reached
quite different levels of internal dynamical evolution. While the age
of people is easily readable in the identity card, determining their
physical shape is not straightforward (and it depends on the capacity
of correctly reading a few characteristics impressed on their
body). The same holds for star clusters. Thus, a proper
characterization of any GC requires the knowledge, not only of its
internal structure and kinematics, but also of its dynamical age.

\subsection{Blue Stragglers as  gravitational test particles of GC dynamics}
The internal dynamical activity of GCs is thought to also generate a
variety of stellar exotica, as blue straggler stars (BSSs) and
interacting binaries containing heavily degenerate objects, like black
holes and neutron stars (see \citealp{bailyn95,ransom+05, cool+98,
  strader+12, cadelano+17, cadelano+18, ema+14, pallanca+13,
  pallanca+17, ferraro+15}). Among these, BSSs are certainly the most
abundant and they are the easiest to distinguish from normal stars in
a color-magnitude diagram (CMD), since they define a sort of sequence
extending brighter and bluer than the cluster MS-TO point, mimicking a
sub-population of young stars (see Fig. \ref{fig:1};
\citealp{sandage53, ferraro+92, ferraro+93, ferraro+97a, ferraro+03,
  leigh07, moretti08, simunovic16, parada+16}).

\begin{figure*}
\includegraphics[scale=0.8]{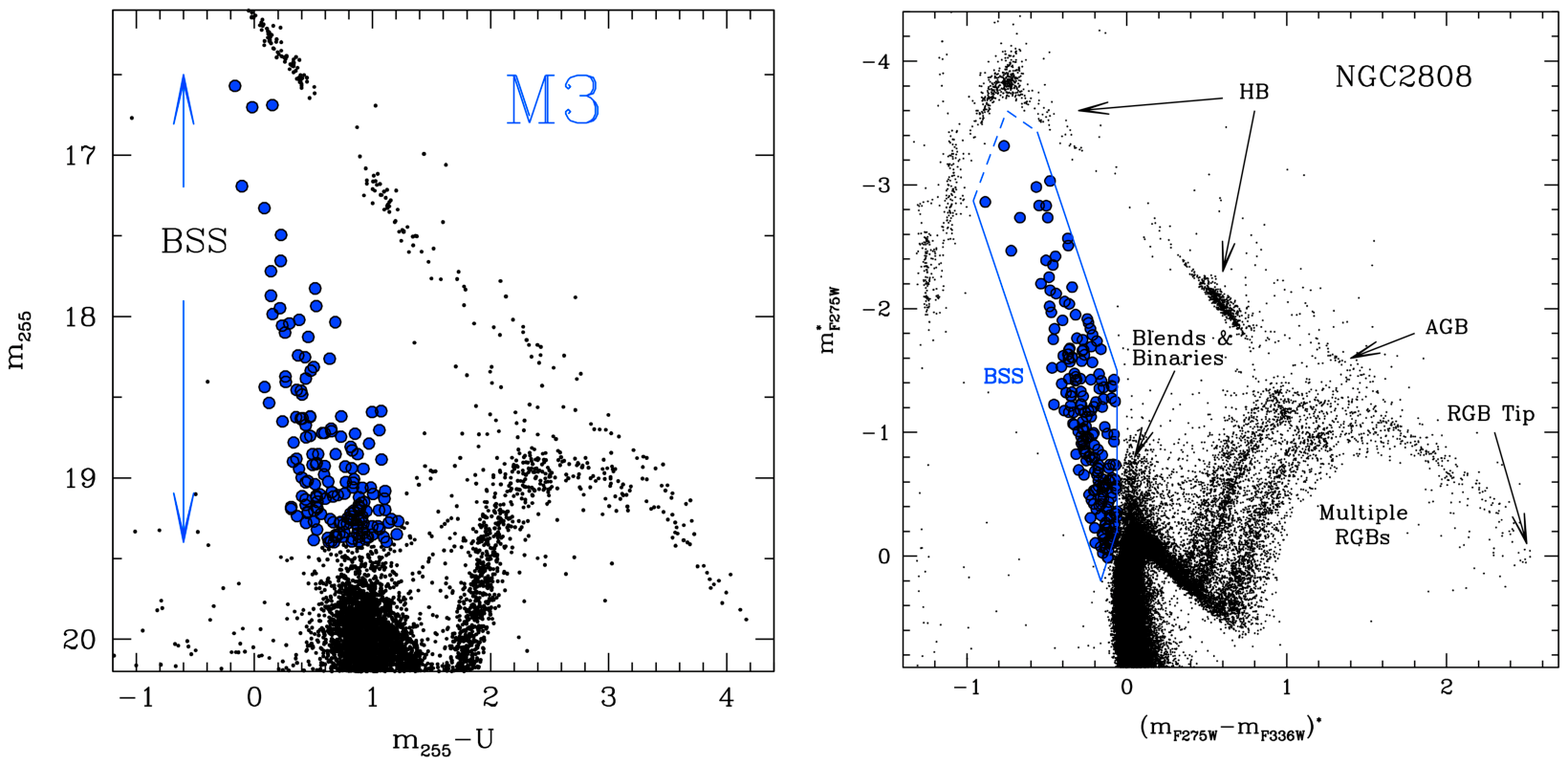}
\caption{Ultraviolet CMDs of two massive globular clusters (namely, M3
  and NGC 2808). In both cases, BSSs are plotted as large blue
  circles. In the right panel, the position of the main evolutionary
  sequences are also marked.}
\label{fig:2}       
\end{figure*}

Their history dates back to 1953, when the American astronomer Allan
Sandage first discovered this puzzling population of stars that seemed
to go against the rules of the stellar evolution theory, in the
Galactic GC Messier 3 (M3). He dubbed them ``stragglers'' because they
are located outside the main evolutionary sequences in the CMD and
they seem to be trailing in the evolution with respect to the vast
majority of stars in the cluster. The presence of these (apparently)
young stars in a GC was completely unexpected, since star formation
essentially stopped 13 billion years ago in these systems. Indeed,
BSSs are not thought to be a young stellar population, and it is
widely accepted that they are hydrogen-burning stars more massive than
the MS-TO objects \citep{shara97,gilliland98}. However, the details of
their formation mechanism are not completely understood yet, and two
main mechanisms are commonly advocated to explain their origin (see
Fig. \ref{fig:1}): (1) direct collisions (COLL), in which the stars
might actually merge, mix their nuclear fuel and ``re-stoke'' the
fires of nuclear fusion \citep{hills76}, and (2) mass-transfer (MT) in
tight binary systems, where the less massive object acts as a
``vampire'', siphoning fresh hydrogen from its more massive companion,
possibly up to the complete coalescence of the two stars \citep{mc64}.
Both processes are suggested to add fresh fuel (hydrogen) into the
stellar core, thus prolonging the star lifetime and making it look
more youthful (blueness and brightness being the attributes of stellar
youth).  Both these processes have an efficiency that depends on the
local environment \citep{fusi92, ferraro+95, ferraro+99, ferraro+06a,
  piotto+04, davies04, ema08a, sollima08, knigge09, chen+09,
  beccari+13, chatterjee+13}, and they can act simultaneously within
the same cluster \citep[e.g.,][]{ferraro+06b, ferraro+09,
  dalessandro+13a, leigh+13, xin15, beccari+19, simon19}.

\begin{figure*}
\includegraphics[scale=0.9]{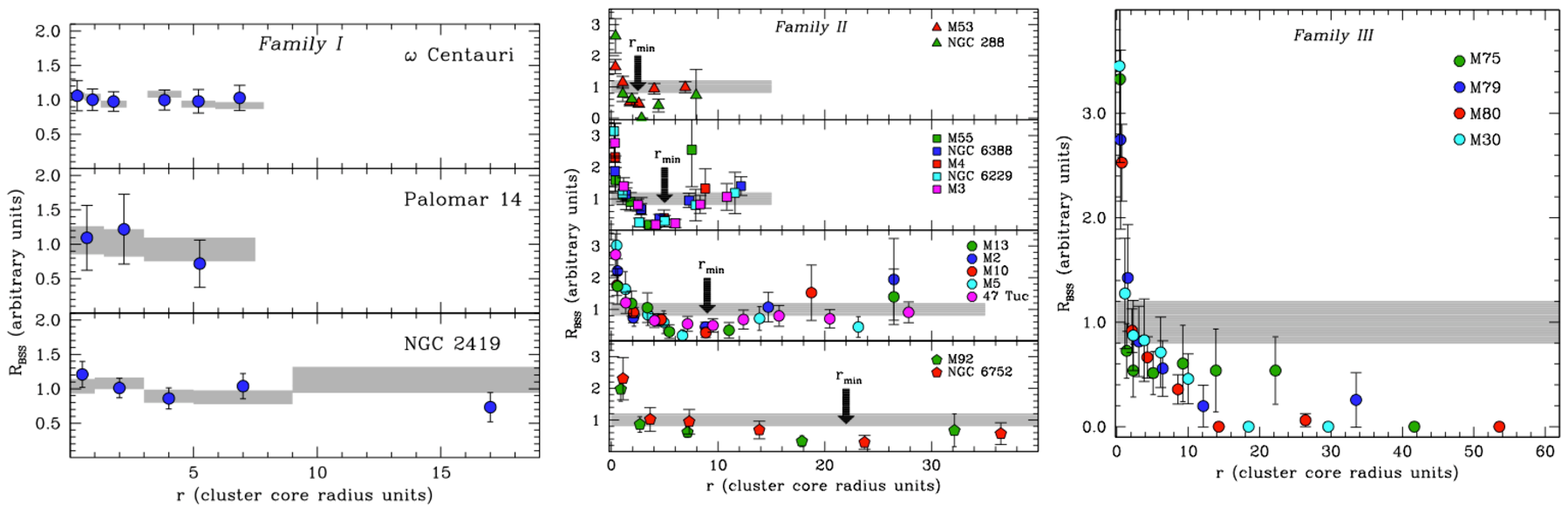}
\caption{Normalized radial distribution of BSSs (colored symbols)
  compared to that of normal cluster stars taken as reference (grey
  strips) in the three main families defined by
  \citet{ferraro+12}:{\it Family I} = dynamically young clusters (left
  panel), {\it Family II} = dynamically intermediate-age clusters
  (central panel),{\it Family III} = dynamically old clusters (right
  panel). }
\label{fig:3}       
\end{figure*}

Irrespective of their formation mechanism, BSSs represent a population
of heavy objects ($M_{\rm BSS}=1.0-1.4 M_\odot$; e.g.,
\citealp{fiorentino+14,raso+19}; see also \citealp{ferraro+16})
orbiting in an ``ocean'' of lighter stars (the average stellar mass in
an old GC is $\langle M\rangle= 0.3 M_\odot$; e.g.,
\citealp{djorg93}). For this reason, BSSs can be used as powerful
gravitational probes to investigate key physical processes (such as
mass segregation and dynamical friction) characterizing the dynamical
evolution of star clusters (e.g., \citealp{ferraro+09, ferraro+12,
  dalessandro+13a, simunovic14}).  {\it But, how difficult is to
  collect complete samples of BSSs in GCs?}

\subsection{The UV route to search for BSSs in GCs}
Because of their high surface temperatures ($T_{\rm eff}\sim
6500$-9000 K), BSSs are among the brightest objects at ultraviolet
(UV) wavelengths in Galactic GCs (see Fig. \ref{fig:2}). Thus, their
systematic search strongly benefitted from the advent of UV space
facilities. More than 20 years ago we promoted the so-called {\it UV
  route to BSS study in GCs} (see \citealp{ferraro+97a, ferraro+97b,
  ferraro+99, ferraro+01, ferraro+03}). This approach consists first
in identifying the stellar sources in images acquired at UV
wavelengths, then, in using the positions of those stars to enable the
source detection in images acquired in the other filters. Such a
technique naturally optimizes the detection of relatively hot stars
and allows the collection of complete sample of BSSs even in the
central region of high-density clusters. Indeed UV CMDs are the ideal
diagrams where to study BSSs, since these stars appear to be clearly
distinguished from the other evolutionary sequences and can be safely
selected (see Fig. \ref{fig:2}).  In particular, (1) together with the
hottest horizontal branch (HB) stars, BSSs are the brightest objects
in UV CMDs, while most of the RGB stars are significantly fainter (at
odds with what happens in the optical diagrams), and (2) BSSs draw a
narrow and well-defined sequence spanning approximately 3 magnitudes.

By following this approach, we derived complete samples of BSSs in the
central cores of several Galactic GCs, including systems of very high
central density (see \citealt{lanzoni+07a, lanzoni+07b, lanzoni+07c,
  ema08b, dalessandro+09, sanna+12, sanna+14, contreras+12,
  dalessandro+13a, dalessandro+13b}).  The dataset recently acquired
within the HST UV Legacy Survey of Galactic Globular Clusters
\citep{piotto+15} allows the extension of this approach to a
significant number of additional clusters (see Section
\ref{sec_apiu}). By using this dataset, \citet{raso+17} quantitatively
demonstrated the clear advantages of the {\it UV-guided} search for
BSSs, with respect to the {\it optical-guided} approach. In fact, the
detailed comparison between the catalogs obtained through the two
different methodologies in four GCs (namely, NGC 2808, NGC 6388, NGC
6541 and NGC 7078) has shown that a large sample of stars in the
innermost region of these systems are missed in the {\it
  optical-guided} case. The number of missed stars depends on the
cluster structure, varying from a few hundreds up to thousands in high
density clusters. The vast majority ($> 70\%$) of the missed stars is
located within the innermost 20''-30" from the cluster centre, thus
demonstrating the potential risk of using optical-driven catalogs to
study the radial distributions and population ratios of BSSs.
   
\begin{figure*}
\includegraphics[scale=0.7]{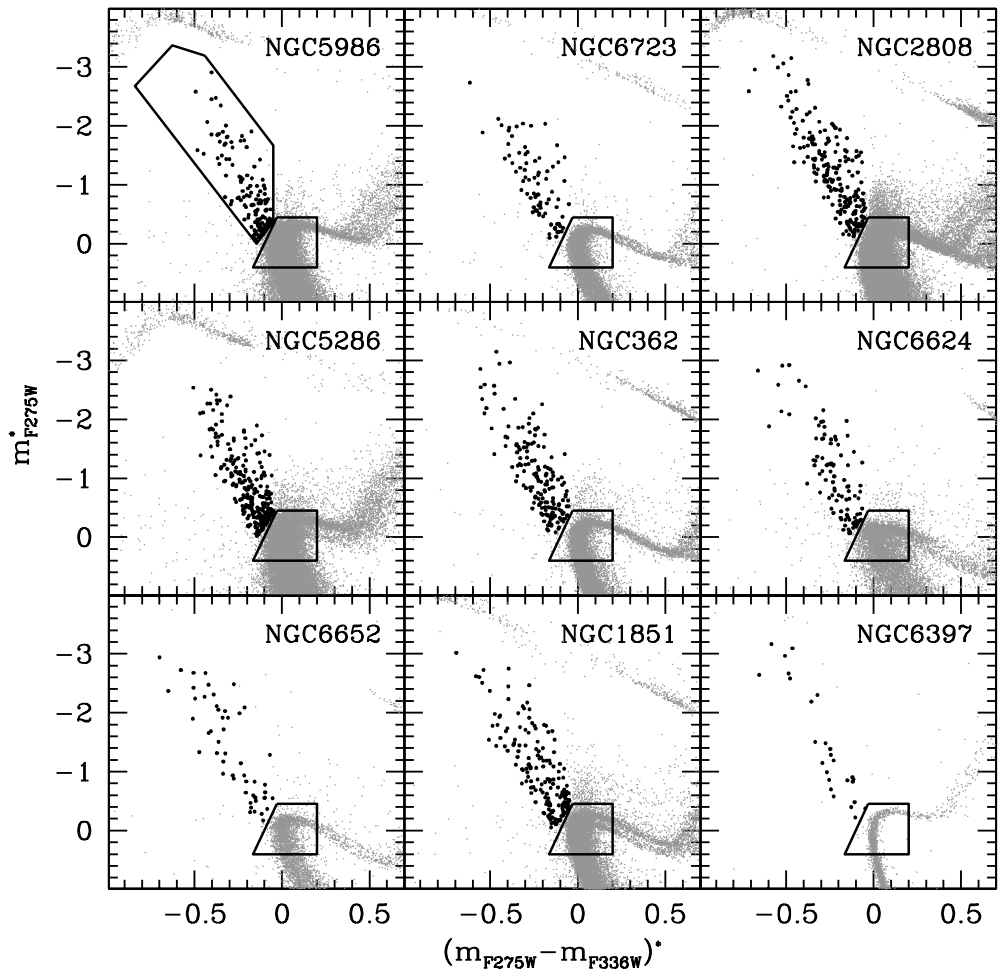}
\caption{The sample of BSSs (black dots) identified in the n-CMD for a
  representative sample of 9 clusters discussed in \citet{ferraro+18}.
  The BSS selection box is drawn in the first panel, the one adopted
  for the MS-TO population is marked for all clusters.}
\label{fig:4}       
\end{figure*}

\section{Setting the ``dynamical clock''}
\label{sec_clock}
According to their formation mechanisms, BSSs are significantly
heavier than the average cluster population. Hence, as the dynamical
evolution of the parent cluster advances, these objects progressively
tend to migrate to the innermost regions. Of course, this modifies the
radial distribution of BSSs, by producing a progressive depletion of
these stars in the outer regions and an increase of their density
toward the cluster center. Hence, the dynamical age of a star cluster
is mirrored by the shape of the radial distribution and by the level
of central sedimentation of its BSS population, exactly as the
physical shape of people is imprinted on their body through many
observable features. Because the progressive flow of BSSs toward the
center measures the dynamical age of the parent cluster in a similar
way as the progressive sedimentation of sand grains in an hourglass
measures the flow of time, we named this method the ``dynamical
clock''.

\subsection{Reading the signature of dynamical evolution from the BSS radial distribution}
\citet{ferraro+12} analyzed the BSS distribution over the entire
radial extension in a sample of 21 Galactic GCs (\citealp{ferraro+97a,
  ferraro+04, ferraro+06a, sabbi+04, lanzoni+07a, lanzoni+07b,
  lanzoni+07c, ema08a, ema08b, dalessandro+09, beccari+06a,
  beccari+06b, beccari+11, contreras+12, sanna+12}) and first
demonstrated that its shape can be used to measure the level of
dynamical evolution reached by the host system.

To this aim, \citet{ferraro+12} used the ``BSS normalized radial
distribution'' (hereafter BSS-nRD), defined as the ratio ($R_{\rm
  BSS}$) between the fraction of BSSs sampled in a radial bin and the
fraction of cluster light sampled in the same bin \citep{ferraro+93}.
Since the number of stars scales as the sampled luminosity, this ratio
is equal to one for any population not affected by dynamical
evolution.  Hence, this parameter is particularly powerful in
quantifying any excess or deficit of stars with respect to the
``normal'' radial distribution, as indeed observed for exotic
populations, like BSSs. Thus, by analyzing the shape of the BSS-nRD,
the surveyed sample of chronologically old and coeval GCs has been
partitioned in three main families (see Fig. Fig. \ref{fig:3}):

\begin{figure*}
\includegraphics[scale=0.7]{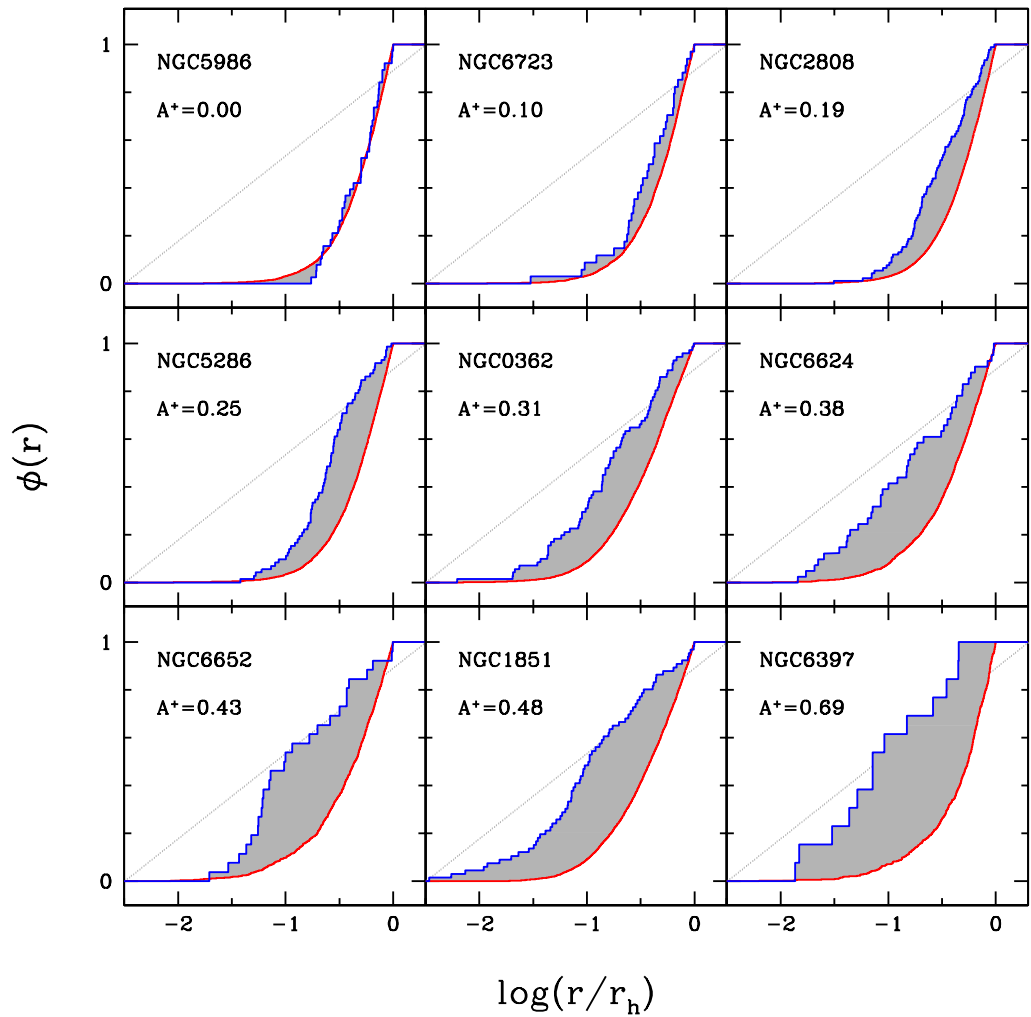}
\caption{Cumulative radial distributions of BSSs (blue line) and REF
  stars (red line) in the nine GCs shown in Figure \ref{fig:4}. The
  horizontal axis provides the logarithm of the cluster-centric
  distance, in units of the half-mass radius $r_h$. The size of the
  area between the two curves (shaded in grey) corresponds to the
  labelled value of $A^+$. Clusters are ranked in terms of increasing
  value of $A^+$.}
\label{fig:5}       
\end{figure*}

\begin{itemize}
\item Family I, where the BSS-nRD is flat (i.e., the radial
  distribution of BSSs within the cluster is fully compatible with
  that of the sampled light and indistinguishable from that of
  ``normal'' stars);
\item Family II, where the BSS-nRD is bimodal, with a high peak in the
  cluster center, a dip at an intermediate radius ($r_{\rm min}$), and
  a rising branch in the external regions (this behavior indicates an
  excess of BSSs in the center and a depletion at intermediate radii,
  with respect to normal cluster stars);
\item Family III, where the BSS-nRD shows only a central peak,
  followed by a monotonically decreasing trend (this indicates a huge
  excess of BSSs in the center and a severe lack of them anywhere else
  in the cluster).
\end{itemize}
This variety of shapes (also detected in extra-Galactic GCs; see
\citealp{li+13}) has been interpreted as the manifestation of the
effect of dynamical friction, which drives the objects more massive
than the average toward the cluster centre \citep[e.g.][]{mapelli+04,
  mapelli+06, miocchi+15}, with an efficiency that mainly depends on
the local star density (i.e., it decreases at increasing radial
distance; see \citealp{alessandrini+14}). Hence, a flat BSS-nRD
indicates that dynamical friction has not affected the BSS population
yet (not even in the innermost regions), and therefore Family I
globular clusters are ``dynamically young'' (left panels in
Fig. \ref{fig:3}). In more evolved GCs (Family II), dynamical friction
has progressively removed BSSs at increasingly larger distances from
the center, thus generating a minimum in the BSS-nRD at increasingly
larger values of $r_{\rm min}$ (from top to bottom, in the central
panels of Fig. \ref{fig:3}).  In Family III systems, dynamical
friction already affected also the most remote BSSs, accumulating all
of these stars toward the cluster center and thus producing a
monotonic BSS-nRD with a central prominent peak; these are
``dynamically old'' GCs (right panel in Fig. \ref{fig:3}).

\subsection{Reading the signature of the dynamical evolution from the BSS segregation level}
\label{sec_apiu}
In \citet{alessandrini+16} we proposed a new parameter ($A^+$) to
measure the level of BSS central sedimentation.  $A^+$ is defined as
the area enclosed between the cumulative radial distribution of BSSs
and that of a reference (REF), lighter, population (as the HB, RGB, or
MS-TO stars), measured within a given distance from the cluster center
(the half-mass radius).  N-body simulations demonstrate that this
parameter systematically increases as a function of time, following
the dynamical evolution of the cluster and, more specifically,
tracking the process of BSS segregation (see Figure 5 in
\citealp{alessandrini+16}).  At initial times, when all stars are
spatially mixed regardless of their mass, BSSs and the REF population
share the same cumulative radial distributions and $A^+=0$. As the
action of dynamical friction proceeds, BSSs migrate toward the centre
of the system and the two curves start to separate, thus providing a
progressively increasing value of $A^+$. \citet{lanzoni+16} measured
the new parameter within one half-mass radius ($A^+_{rh}$) for the
same set of GCs discussed in \citet{ferraro+12}, finding a tight
correlation with $r_{\rm min}$ (see their Figure 2). This demonstrates
that both parameters measure the effect of dynamical friction: as
clusters get dynamically older, dynamical friction progressively
removes BSSs at increasingly larger distances from the center (thus
generating a minimum at increasingly larger values of $r_{\rm min}$)
and accumulates them toward the cluster center (thus increasing
$A^+$). In addition, \citet{lanzoni+16} found a strong correlation
between $A^+$ and the central relaxation time of the cluster
($t_{rc}$) (similar to that found by \citealp{ferraro+12} using the
parameter $r_{\rm min}$), thus fully confirming that $A^+$ can be
efficiently used to measure the level of dynamical evolution reached
by star clusters (see also the simulations by \citealp{sollima+19}).
 
\begin{figure}
\includegraphics[scale=0.4]{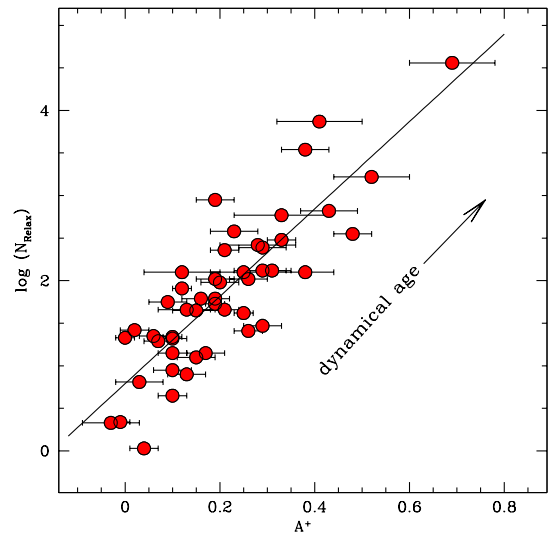}
\caption{The strong correlation between $A^+ $ and $\log(N_{\rm
    relax})$ for the sample of 48 GCs discussed in \citet{ferraro+18}.
  The parameter $N_{\rm relax}$ quantifies the number of current
  central relaxation times occurred since cluster formation. The tight
  relation between these two parameters demonstrates that the
  segregation level of BSSs measured by $A^+$ can be used to evaluate
  the level of dynamical evolution experienced by the parent
  cluster. The best fit relation (eq. 1) is also shown as a solid
  line. The arrow indicates increasing dynamical ages.}
\label{fig:6}       
\end{figure}

This approach was recently extended \citep{ferraro+18} to 27
additional systems observed within the HST UV Legacy Survey of
Galactic Globular Clusters (\citealp{piotto+15}, see also
\citealp{raso+17}). Combined with the clusters studied in
\citet{lanzoni+16}, this provided us with a total sample of 48 GCs,
corresponding to almost $33\%$ of the entire Milky Way population.  To
perform a fully homogeneous selection of BSSs in clusters with
different values of distance, reddening and metallicity,
\citet{ferraro+18} made use of ``normalized'' CMDs (see also
\citealp{raso+17}), where the magnitudes and colors of all the
measured stars in a given cluster are arbitrarily shifted to locate
the cluster MS-TO at (0,0) coordinates (see, e.g., Fig. \ref{fig:4},
where the ``normalized'' magnitudes and colors are indicated with
$m_{\rm F275W}^*$ and $(m_{\rm F275W} -m_{\rm F336W})^*$,
respectively). The advantage of using n-CMDs is that, independently of
the cluster properties, BSSs are expected to populate the same region
of the diagram. Hence, the same selection box can be used in all GCs
for a homogeneous selection of the BSS population. For the sake of
illustration, Fig. \ref{fig:4} shows a sample of nine n-CMDs analyzed
in \citet{ferraro+18}, with the BSS selection box marked in the
top-left panel.

To compute the $A^+$ parameter, the radial distribution of BSSs must
be compared with that of a REF population of normal cluster stars
tracing the overall density profile of the system. As REF population,
\citet{ferraro+18} adopted MS stars in the MS-TO region, since this
portion of the CMD includes several hundred objects and therefore is
negligibly affected by statistical fluctuations.  Fig. \ref{fig:5}
illustrates the cumulative radial distributions obtained for the
sub-sample of nine clusters shown in Fig. \ref{fig:4}, covering the
entire range of values measured for $A^+$ in the full sample of 48 GCs
(NGC 5986 the lowest value, and NGC 6397 having the largest one).

To investigate the connection between the BSS segregation level and
the dynamical status of the parent cluster, we studied the relation
between the measured values of $A^+$ and the number of current central
relaxation times ($t_{rc}$) that have occurred since the epoch of
cluster formation ($t_{\rm GC}$): $N_{\rm relax} = t_{\rm
  GC}/t_{rc}$. Because all Galactic GCs have approximately the same
age, for all the program clusters we assumed $t_{\rm GC} = 12$ Gyr
(see the compilation of \citealp{forbes+10}).  A strong correlation
was found (Fig. \ref{fig:6}):
\begin{equation}
  \log N_{\rm relax} = 5.1 \times A^+ +0.79,
\end{equation}
clearly demonstrating that $A^+$ is a powerful indicator of GC
internal dynamical evolution. This is indeed a key relation: it allows
the empirical determination of the dynamical age of a star cluster
from just the measure of the central sedimentation level of its BSS
population. By using this relation it is possible to learn how much
dynamically-old is a GC, and if it is more or less evolved than other
systems with comparable or different structural/dynamical properties.

\begin{figure}
\includegraphics[scale=0.7]{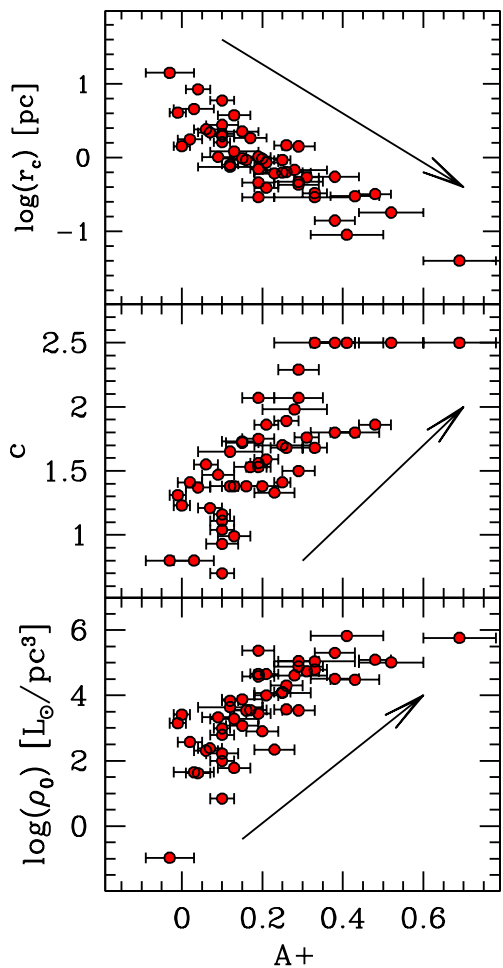}
\caption{Relation between $A^+$ and three physical parameters that are
  expected to change with the long-term dynamical evolution of GCs:
  the core radius $r_c$ ({\it top panel}, the concentration parameter
  c ({\it middle panel}), and the central luminosity density $\rho_0$
  ({\it bottom panel}).  The values of $r_c$, c and $\rho_0$ are taken
  from \citet{harris96}. The arrows indicate increasing dynamical
  ages.}
\label{fig:7}       
\end{figure}

Measuring the $A^+$ parameter offers the opportunity to empirically
describe the effect of the dynamical aging of star clusters on their
structural parameters.  Fig. \ref{fig:7} shows the behavior of the
core radius ($r_c$), the concentration parameter c (defined as the
logarithm of the ratio between the tidal and the core radii), and the
central luminosity density ($\rho_0$; all are taken from
\citealt{harris96}, 2010 edition), as a function of $A^+$.  Quite
well-defined trends are apparent in the figure, with $r_c$ decreasing,
and c and $\rho_0$ increasing with $A^+$ (i.e., with increasing
dynamical age), confirming that star clusters tend to develop small
and dense cores as a consequence of their long-term internal dynamical
evolution, in nice agreement with what expected from the theoretical
framework.
  
\begin{figure}
\includegraphics[scale=0.43]{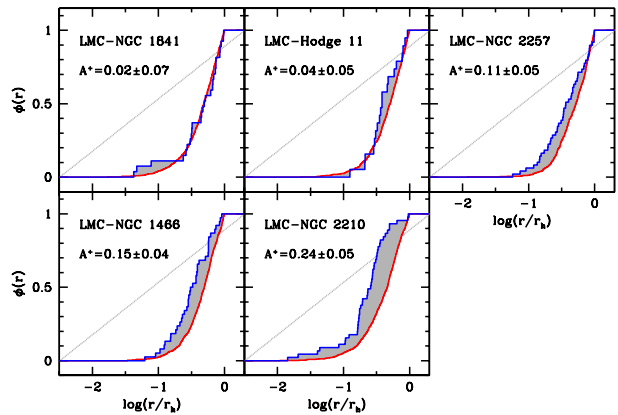}
\caption{Cumulative radial distributions of BSSs (blue line) and REF
  stars (red line) for the five GCs in the LMC discussed by
  \citet{ferraro+19}. The size of the area between the two curves
  (shaded in grey) corresponds to the labelled value of
  $A^+$. Clusters are ranked in terms of increasing value of $A^+$.}
\label{fig:8}       
\end{figure}

\section{The ``dynamical clock'' beyond the Galaxy: an alternative reading of the core size-age enigma in the Large Magellanic Cloud clusters}   
The next obvious step in this line of investigation was to apply the
same method to star clusters in other galaxies, and the Large
Magellanic Cloud (LMC) is indeed the closest, and hence the most
natural, target. Moreover the LMC is also the most intriguing one. In
fact, there is a 30 year old dilemma related to the LMC clusters: the
so-called ``core size-age conundrum'' \citep{mackey+03}. It can be
summarized in just one question: {\it why all the young star clusters
  in the LMC are compact, while the old ones show both small and large
  core radii?}. One of the most commonly accepted solution to this
dilemma was that all clusters in the LMC formed compact, then they
suffered more or less significant expansions of their core driven by
populations of binary black holes \citep{mackey+08}. However, studies
of GC dynamical ageing in the Milky Way show that compact cores tend
to be developed as time passes (top panel in Fig. \ref{fig:7}), which
is just the opposite of what suggested to occur in the LMC.  Hence,
deeper investigations appeared to be necessary to solve the $r_c$-age
dilemma in this external galaxy.  \citet{li+19} applied the
``dynamical clock'' to measure the dynamical evolutionary stage of 7
intermediate-age (between 700 Myr and 7 Gyr) clusters in the LMC,
finding a low-level of dynamical evolution. On the other hand, the
dynamical ageing effects are expected to be most evident in old star
clusters, with chronological ages comparable to those of the Milky Way
systems (12-13 Gyr). Hence, \citet{ferraro+19} selected five 13
Gyr-old GCs in the LMC, showing different core sizes (ranging from 1
to 7 parsec), and measured their stage of internal dynamical evolution
via the BSS sedimentation level.

\begin{figure}
\includegraphics[scale=0.4]{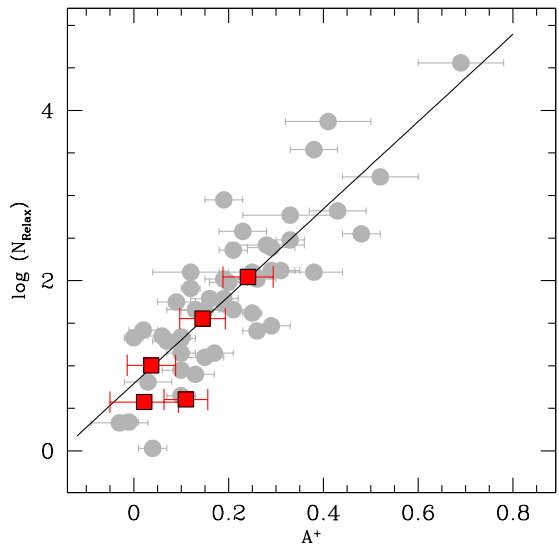}
\caption{The application of the ``dynamical clock'' to extra-Galactic
  clusters: $N_{\rm relax}-A^+$ relation for the five LMC clusters
  discussed in \citet{ferraro+19} (large red squares). For the sake of
  comparison, the sample of 48 Galactic GCs previously investigated
  \citep{ferraro+18} is shown with grey circles. The LMC GCs ranked
  for increasing value of the dynamical age ($A^+$) are NGC 1841,
  Hodge 11, NGC 2257, NGC 1466, and NGC 2210. The arrow indicates
  increasing dynamical age.}
\label{fig:9}       
\end{figure}

\begin{figure}
\includegraphics[scale=0.4]{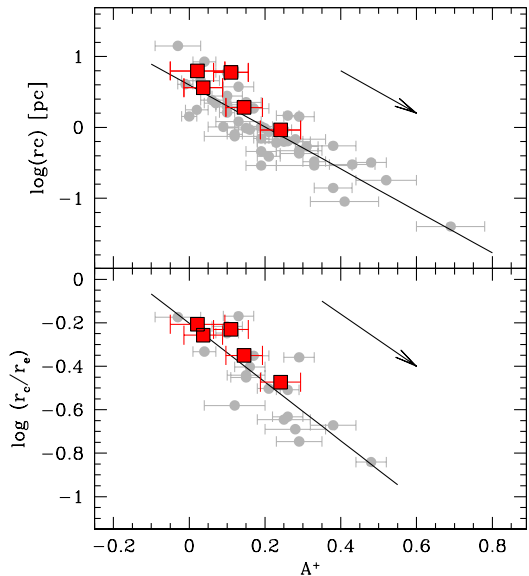}
\caption{{\it Top panel:} the $r_c-A^+$ relation for the five LMC
  clusters discussed in \citet{ferraro+19} (large red squares),
  compared to that obtained for Galactic GCs (grey circles). The arrow
  indicates increasing dynamical age. {\it Bottom panel:} the relation
  between $A^+$ and the ratio between the core radius and the
  effective radius ($r_e$) for the five investigated LMC clusters (red
  squares), compared to that obtained for 19 GCs in the Milky Way
  (grey circles; $r_e$ is from \citealp{miocchi+13}).}
\label{fig:10}       
\end{figure}

The application of the dynamical clock in the LMC clusters is much
more tricky than in our galaxy, because field stars (not belonging to
the clusters) intervening along the line of sight can strongly
contaminate the region of the CMD where BSSs are selected. The
situation appeared particularly critical for two clusters (Hodge 11
and NGC 2210). However the use of appropriate parallel HST
observations in the cluster neighborhoods, allowed us to statistically
decontaminate the BSS samples.  Indeed, the LMC field stars turned out
to be slightly brighter and redder than the bulk of genuine BSSs, thus
affecting only the reddest portion of the population. Hence, the
statistical decontamination was quite effective and the results were
surprisingly stable: over 5000 repeated random subtractions, the
values of $A^+$ showed a very peaked distribution with a small
dispersion, thus certifying the reliability of the measure.

\begin{figure}
\includegraphics[scale=0.4]{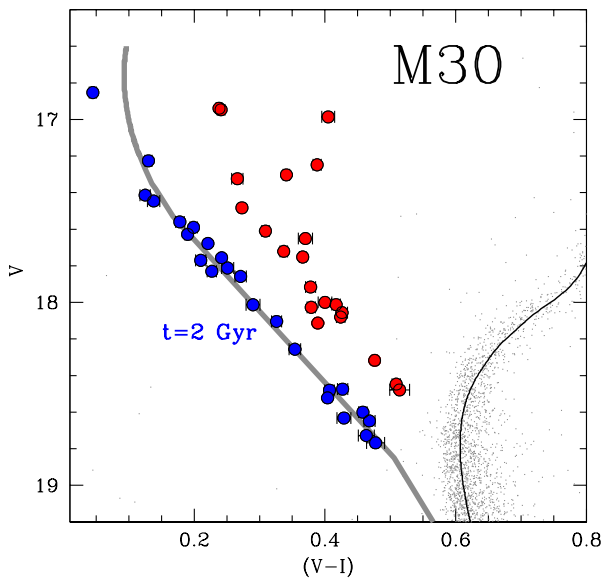}
\caption{The double BSS sequence in M30 (from
  \citealp{ferraro+09}). The thick grey line is the 2 Gyr collisional
  isochrone from \citet{sills09}. The black line is the 12 Gyr-old
  isochrone, best reproducing the cluster MS-TO
  \citep[from][]{cariulo+04}.}
\label{fig:11}       
\end{figure}

The estimate of the BSS sedimentation level in all five selected
clusters confirms that these systems are at different stages of
internal dynamical evolution (see Fig. \ref{fig:8}).  Indeed, $A^+$
shows a nice correlation with the number of relaxation times they
suffered since formation, and the impressive match with the trend
defined by the Galactic population (grey circles) demonstrates that
the ``dynamical clock'' can be efficiently used in any stellar
environment and is weakly affected by the tidal field of the host
galaxy.

The five surveyed LMC GCs also follow the tight correlation between
dynamical age and core radius found by \citet{ferraro+18} for the
Galactic systems (top panel of Fig. \ref{fig:10}), confirming that the
long-term dynamical evolution tends to generate compact objects. This
result has been recently confirmed by \citet{lanzoni+19} who
re-computed all relevant structural parameters of these five LMC
clusters from newly determined star density profiles. In particular,
they re-determined the core radius and the effective radius ($r_e$),
which is defined as the radius of the circle that, in projection,
includes half the total counted stars. The ratio between $r_c$ and
$r_e$ has been found to correlate with the dynamical age in a sample
of 19 Galactic GCs, with systems characterized by large values of
$r_c/r_e$ being dynamically younger than those showing small values of
this ratio \citep[see][]{miocchi+13}, as expected from dynamical
evolution driven by two-body relaxation. The bottom panel of Figure
\ref{fig:10} shows that the 5 LMC GCs nicely fit the trend defined by
the Galactic population, and reveal that no dynamically-old system
with large value of $r_c/r_e$ exists. In turn, this indicates that the
observed properties of these systems are just consistent with their
natural dynamical ageing, with no need of anomalous energy sources
responsible for significant core expansion (as, e.g., a population of
binary black holes; \citealp{mackey+08}).

These results confirm that the internal dynamical evolution tends to
produce clusters with more and more compact cores, and demonstrate
that the spread in core size observed for the old LMC GCs is the
natural consequence of these processes. On the other hand, the
analysis discussed in \citet{ferraro+19} also demonstrates that only
low-mass systems have been recently (in the last $\sim 3$ Gyr) formed
in the LMC, and essentially all of them were generated in the
innermost regions of the galaxy. Hence, they surely cannot resemble
the progenitors of the currently old clusters, which are much more
massive (by up to factors of 100) and orbit at any distance from the
centre of the LMC. Among the young generation of (low-mass) clusters,
only the most compact systems survived and are observable: hence we do
not see young clusters with large core-size just because, if formed,
they have been already disrupted by gravitational interactions within
the LMC potential well. These results suggest a new reading of the
long-standing $r_c$-age distribution of LMC GCs, which does not
require the action of a population of black holes (as suggested by
\citealp{mackey+08}), and is just the natural manifestation of the
long-term internal dynamical evolution \citep{ferraro+19}.  Thus, from
one side, this deepens our understanding of the processes that govern
the internal dynamical evolution of dense stellar systems in different
environments, allowing a direct connection between the old GCs in the
LMC and those in our Galaxy. From the other side, these findings offer
an alternative (and less exotic) reading of the long-standing LMC
conundrum, with a strong impact on our understanding of star cluster
formation and their evolution over cosmic time.  Moreover, as it often
happens in science, while answering an old dilemma, this discovery
poses a new, probably deeper, question that needs to be addressed:
{\it why did only relative low-mass clusters form in the last 3 Gyr in
  the LMC?}

\begin{figure*}
\includegraphics[scale=0.7]{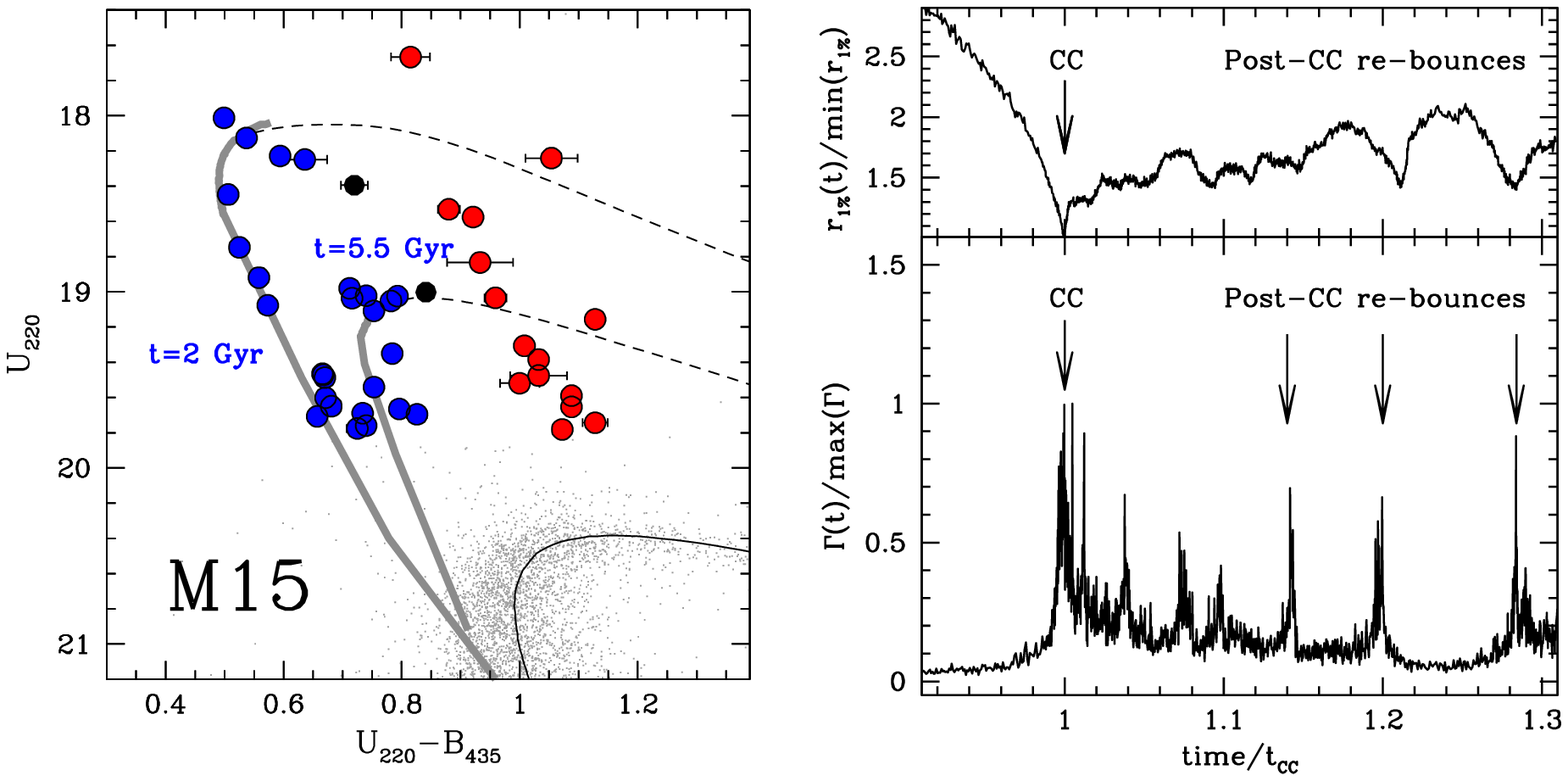}
\caption{{\it Left Panel:} The double BSS sequence in M15 (from
  \citealp{beccari+19}). Two branches are apparent along the blue
  sequence: they are reproduced by a 2 Gyr-old and a 5.5 Gyr-old
  collisional isochrone, respectively. {\it Right Panel:} The
  evolution in time of the $1\%$ Lagrangian radius (top panel) and of
  the collisional parameter (bottom panel) in dynamical simulations.
  The epochs of core collapse and the main post-core collapse
  re-bounces are indicated with arrows.}
\label{fig:12}       
\end{figure*}

\section{Exploring the post-core collapse (PCC) phase}
\label{sec_CC}
The BSS observational features might also provide crucial information
about the most spectacular dynamical event in cluster lifetime: core
collapse (CC). This was first realized by \citet{ferraro+09} with the
discovery of two well-separated and almost parallel sequences of BSSs
in the post-core collapse (PCC) cluster M30 (Fig. \ref{fig:11}). This
was the very first time that such a feature was detected in any
stellar system, opening the possibility to disentangle between the two
BSS formation processes. In fact, the comparison with evolutionary
models of BSSs formed by direct collisions of two MS stars
\citep{sills09} showed that the blue-BSS sequence is well reproduced
by a collisional isochrone with an age of $\sim 2$ Gyr.  Instead, the
red-BSS population is far too red to be reproduced by collisional
isochrones of any age, and it turned out to be located in the portion
of the CMD where MT binary models are expected to lie
(\citealp{tian+06, xin15}; see also recent models by
\citealp{jiang+17} showing that MT-BSSs might also ``contaminate'' the
blue-BSS sequence).  The impressive agreement between the blue-BSS
sequence and the collisional isochrone suggests that the vast majority
of these stars have a collisional origin.  Due to standard stellar
evolution, they are destined to move toward the red in the CMD, in
typical timescales of a few Gyrs that are shorter for brighter (more
massive) stars. Hence, the extension in magnitude of the blue sequence
and the existence of a clear-cut gap between the two chains suggest
that these BSSs are nearly coeval and have been generated by a {\it
  recent} and {\it short-lived} event. This latter characteristic is
typical of CC \citep{djo+86}, during which the stellar collision rate
is also known to significantly increase. On the basis of these
considerations, \citet{ferraro+09} concluded that {\it most BSSs along
  the blue sequence formed simultaneously during the CC event,
  approximately 2 Gyr ago.}  This scenario has been fully confirmed by
\citet{simon19}, who presented detailed stellar merger simulations for
M30, concluding that the BSS distribution along the blue sequence is
well consistent with a burst of formation started $\sim 3.2$ Gyr ago,
a duration of approximately 0.93 Gyr and a peak of formation rate of
30 BSSs/Gyr. Such an impressive formation rate was possibly triggered
by the cluster CC. Instead, the BSSs along the red sequence formed
with a much more modest and constant rate of $\sim 2.8$ BSSs/Gyr over
the last 10 Gyr.

Interestingly, the double BSS feature has been later detected in two
additional clusters: NGC 362 \citep{dalessandro+13a} and NGC 1261
\citep{simunovic14}. The case of NGC 2173, an intermediate-age cluster
in the LMC, is instead still debated, since the feature can possibly
be due to the contamination of the BSS population from LMC field stars
(see \citealp{li+18a, li+18b, ema+19a, ema+19b}).

{\it M15: another surprise --} By applying an advanced photometric
de-blending technique to a set of high-resolution images,
\citet{beccari+19} discovered an even more peculiar double BSS
sequence in the innermost regions of the PCC cluster M15 (see
Fig. \ref{fig:12}). Also in this case the red BSS sequence cannot be
reproduced by collisional isochrones of any age, but this time the
blue BSS sequence showed a quite complex structure. In fact two
distinct branches are visible: the first branch appears to be
extremely narrow and it extends up to 2.5 mag brighter than the
cluster MS-TO point, while the second branch extends up to 1.5 mag
from the MS-TO.  The comparison with formation models of collisional
BSSs \citep{sills09} indicates that both these populations formed
through this channel, at two different epochs: approximately 5.5 and 2
Gyr ago, respectively. The two branches could therefore be the
observational signatures of two major collisional episodes suffered by
M15, likely connected to the most advanced stages of dynamical
evolution of its core: the first one (possibly tracing the beginning
of CC) occurred approximately 5.5 Gyr ago, while the most recent one
(possibly associated with a core oscillation in the PCC evolution)
dates back to 2 Gyr ago. This scenario is consistent with the results
of Monte Carlo simulations (Fig. \ref{fig:12}), showing that the deep
initial CC (clearly distinguishable at $t/t_{\rm CC} = 1$) leads to
the largest increase of the collision rate (${\rm \Gamma}$), and it is
followed by several distinct re-collapse episodes leading to secondary
peaks of the ${\rm \Gamma}$ parameter.

These results further provide strong evidence in support to the tight
connection between the BSS properties and the internal dynamical
evolution of collisional stellar systems, also opening new
perspectives in the study of the most extreme phenomena, as the CC
event and post-CC evolution.

\begin{acknowledgements}
This paper is part of the project Cosmic-Lab (``Globular Clusters as
Cosmic Laboratories'') at the Physics and Astronomy Department of the
Bologna University (see the web page:
http//www.cosmic-lab.eu/Cosmic-Lab/Home.html).  The authours
acknowledge the Accademia dei Lincei for offering the opportunity to
write this contribution. The authors also aknowledge the anonymous
referee for the careful reading of the manuscript.
\end{acknowledgements}

\section*{Funding}
The research is funded by the project Dark-on-Light granted by MIUR
through PRIN2017 contract (PI: Ferraro).

\section*{Compliance with ethical standards}
The manuscript complies to the Ethical Rules applicable for this
journal.

%
\section*{Conflict of interest}
The authors declare that they have no conflict of interest.



\end{document}